# Research Collaborations Within and Across Disciplines in NIH-Funded Atlas Programs

Michael P. Ginda[1], Yongxin Kong[1], Katy Börner[1,*]

[1] Department of Engineering, Luddy School of Informatics, Computing, and Engineering, Indiana University, Bloomington, IN, USA

* Corresponding authors
katy@iu.edu

## Abstract
The National Institutes of Health (NIH) fund several programs (aka consortia) that support the construction of organ-specific or whole person atlases. These initiatives are designed to build comprehensive open reference datasets while advancing knowledge graphs, analytics, visualizations, and user interfaces needed to explore multiscale human atlas data. Here, we compare the Human Tumor Atlas Network (HTAN), Human BioMolecular Atlas Program (HuBMAP) and Cellular Senescence Network (SenNet) to examine collaboration by consortium component (CC) and researcher role and its relation to scientific impact. We retrieved 884 NIH award records and identified 1,255 focal publications acknowledging these awards, involving 8,952 researchers; compiled 99,108 citing publications covering 93% of the focal publications; constructed co-authorship networks; and as single-project publications or internal collaborations (IC), including intra-CC and inter-CCs collaborations. within each atlas program. Results show that IC publications are rare but involve many funded projects and principal investigators (PIs). Technology-oriented CCs tend to engage in intra-CC collaboration, whereas experimental-oriented CCs often collaborate across CCs of the same type. Compared with other researchers, PIs exhibited a significantly stronger association with inter-CCs than with single-project publications. IC publications have higher mean Relative Citation Ratios than single-project publications. Inter-CCs publications have significantly higher cumulative and annual citation levels in the first five post-publication years. Study methods and results can be used to inform the funding, management, and impact reporting for future Big Science programs that bring together 300+ experts from different disciplines to address major research challenges.

## Introduction
Biomedical atlas programs fund highly distributed teams to generate experimental data, open source code, and novel reference maps of the human body (Börner et al. 2025; de Bruijn et al. 2025; Börner et al. 2026). Team science research studies the impact of organizational structures and processes on the success of Big Science collaborations (Börner et al. 2021; Hall et al. 2018; Vogel et al. 2021). Understanding collaboration in biomedical atlas programs therefore requires examining how their functional organization relates to collaborative activity and scholarly outcomes.

Over the last eight years, the National Institutes of Health (NIH) has supported 96 projects within three major atlas-building programs. These programs have distinct scientific goals:  The Human BioMolecular Atlas Program (HuBMAP) maps the healthy human body at cellular resolution (HuBMAP Consortium et al. 2019; Jain et al. 2023). The Cellular Senescence Network (SenNet) characterizes senescent cells across tissues and the lifespan (SenNet Consortium et al. 2022; Börner et al. 2026). The Human Tumor Atlas Network (HTAN) maps tumor development and progression (de Bruijn et al. 2025; Rozenblatt-Rosen et al. 2020). Collaborative work within these programs is distributed across funded projects that are grouped into 'consortium components' (CCs) with different funding, functions, and organizational arrangements (**Table 1**). Consortium component types make it possible to compare collaboration within and across functional boundaries.

**Table 1. Comparison of NIH-funded atlas programs consortium components.**

| Component Type | Human Tumor Atlas Network (HTAN) | Human Biomolecular Atlas Program (HuBMAP) | Cellular Senescence Network (SenNet) |
|---|---|---|---|
| **Data Coordination** Centers | Data Coordinating Center (DCC) | HuBMAP Integration, Visualization and Integration (HIVE) Collaboratory | Consortium Organization and Data Coordination Center (CODCC) |
| **Experimental** Research Centers | Pre-Caner Atlas (PCA)<br><br>Human Tumor Atlas (HTA) | Tissue Mapping Centers (TMC H)<br>Demonstration Projects (DEMO) | Tissue Mapping Centers (TMC S) |
| **Technology** Development Centers | | Transformative Technology Development (TTD)<br>Rapid Technology Integration (RTI) | Technology Development and Application (TDA) |

Science of science approaches provide methods and tools to examine programs through scholarly and administrative records (Fortunato et al. 2018). Comparative analyses of Big Science projects have combined funding, publication, citation, and coauthorship data to characterize program development over time (Börner et al. 2021). Project-level evaluations have likewise linked funding-agency and bibliographic metadata to examine research productivity and influence (Perianes-Rodríguez et al. 2024). In the atlas construction context, studies of the Human Reference Atlas effort have connected publications, experts, funded projects, and experimental datasets to characterize its evidence base in support of atlas construction and use (Kong et al. 2024; Kong and Börner 2024).

Research collaboration has been classified across multiple organizational, geographic, and disciplinary levels. At the organizational level, collaboration is commonly distinguished within and across research groups (Iglič et al. 2017), departments (Pepe and Rodriguez 2010), and institutions (Fan et al. 2020), with further distinctions based on institutional characteristics, such as academic, governmental, and private-sector affiliations (Boardman and Corley 2008). At the geographic level, collaboration is differentiated by spatial proximity and national boundaries, distinguishing domestic and international partnerships (Abramo et al. 2022). Disciplinary classifications distinguish collaboration within and across scientific fields, using researchers' disciplinary affiliations or publication subject classifications (Basner et al. 2013; Arhiliuc and Guns 2023). Within large research initiatives, program membership and funding relationships provide a more specific basis for identifying collaboration patterns. Program-based classifications distinguish collaboration within and across research programs (Fagan et al. 2018), whereas funding-based approaches identify publications supported by single or multiple funding hubs (Llewellyn et al. 2020) or partnerships among organizations participating in funded research projects (Smith et al. 2023).

At the researcher level, research collaboration has been examined in relation to academic rank, career stage, and project roles. Academic rank has been associated with differences in collaboration preferences, with lower-ranked academics favoring within-institution collaboration and higher-ranked academics showing a greater tendency toward international collaboration (Abramo et al. 2014). Across career stages, researchers' roles within scientific teams evolve, accompanied by differences in collaboration frequency and the number of collaborators (Lu et al. 2021). Within large research initiatives, project roles have been associated with differences in collaboration patterns. PIs have been found to engage in more extensive collaboration, including international and interdisciplinary research, than the broader researcher population (Kastrin et al. 2018). In multicenter research networks, center leaders were more likely to establish ties with researchers in other roles (The Centers for Population Health and Health Disparities Evaluation Working Group and Okamoto 2015), while PIs maintained more collaborative ties than non-PI members and played a greater role in connecting researchers across centers (Ducharme et al. 2024). Related research has also found that non-leadership members can bridge different research groups (Punjabi et al. 2026).

Citation impact is widely used to assess scientific influence and provides an important basis for evaluating research collaboration (Hutchins et al. 2016; Bu et al. 2021). Studies have compared citation impact using normalized citation indicators, post-publication citation windows, and the origins of citing publications, showing that citation advantages vary across collaboration patterns and research contexts (Yegros-Yegros et al. 2015; Fan et al. 2020; Jue Wang et al. 2024). In related research on interdisciplinary knowledge integration, Larivière et al. (2015) found that papers citing references from more topically distant subdisciplines received higher relative citation counts. Within large funded research initiatives, the relationship between collaboration and citation impact has been examined at different organizational levels. At the grant level, NIH transdisciplinary center grants produced more publications and involved larger author teams than investigator-initiated grants, but had comparable journal impact factors (Hall et al. 2012). Across the NIH-funded publication portfolio, internationally coauthored publications showed higher field-normalized citation impact than publications involving only domestic authors (Grubbs et al. 2019). At the publication level, combinations of national, international, and industry funding in cancer research did not necessarily yield complementary citation benefits (Rotolo et al. 2023), whereas publications supported by multiple NIH Clinical and Translational Science Awards (CTSA) hubs showed greater citation influence than single-hub publications (Llewellyn et al. 2020). Evaluations of individual NIH-funded biomedical research centers have also examined co-authorship structures alongside RCR-based citation influence (Bragg et al. 2022).

However, existing studies have largely examined collaboration patterns, researcher roles, and citation impact at the institutional, research program, or funding-hub level, providing limited insight into how these relationships operate within the functional organization of large research consortia. In particular, how funded projects collaborate within and across CCs with similar or different functions, who participates in these collaborations, and how they relate to citation impact remain insufficiently understood. In this paper, we compare three NIH-funded atlas programs: HuBMAP, SenNet, and HTAN, which focus on generating data across multiple human organs and have complex consortial program structures. We retrieve 884 funding records, 1,355 publications and 8,952 experts from NIH RePORTER, and 99,108 citations from OpenAlex. The resulting multi-modal network links award records to publications and their citations and coauthorships. We classify publication-derived collaborations according to consortium component types of funded projects. An intra-CC collaboration links two or more funded projects within the same CC, whereas an inter-CCs collaboration links funded projects across two or more CC within the same atlas program. We compare award and funding structures, publication productivity, and impact.

There are three research questions in this paper: (1) How do collaboration patterns vary across CC types? (2) How do principal investigators and other researchers differ in their participation, activity, and co-authorship pattern across collaboration patterns? (3) How does publication-level citation impact vary across collaboration patterns?

# Methods

The data preparation and data analysis was primarily completed using the R programming language (R Core Team 2020) and R Studio IDE. Initial R scripts and markdown notebooks were generated using the Claude models Sonnet 4.5 and Haiku 4.6. The scripts and notebooks generated by these models were evaluated and significantly revised to better align with data preparation and analysis requirements. The data, scripts and markdown notebooks that were created for this research are shared via GitHub repository (https://github.com/cns-iu/nih-atlas-supporting-information) for review and re-use by other researchers.

## Data collection and preparation

Data preparation uses a set of sequenced R scripts (R version 4.5.3), each corresponding to a specific data preparation task. Certain data preparation steps are performed outside of the R environment, including: citation data collection in manually from a publication database; name disambiguation is performed using OpenRefine (version 3.10.1) and its suite of string comparison algorithms (Ham, 2013), and network layouts and visualization are created using Gephi (version 0.10.1) network visualization software (Bastian et al., 2009).

### Funding data

Opportunity numbers identify the funding notices from NIH funded projects with specific research programs and identify related sets of awards that were granted under a program. We manually collected sets of opportunity numbers for each NIH atlas program in the study using program administrative sites—specifically, HuBMAP and SenNet programs' funded research sites (National Institutes of Health (NIH) 2024a, 2024b), and programs linked from National Cancer Institute HTAN program site (National Cancer Institute 2026). Importantly, each set of opportunity numbers is linked to the CCs of its respective atlas program and their corresponding CC types (e.g., data coordination, experimental, and technology; see Table 1). Opportunity numbers are how we associate CCs to award data, which are linked to publications.

Award records (**Table 2**) were collected from NIH RePORTER using a two-stage project-identification procedure. First, RePORTER was queried using the funding opportunity numbers identified for each atlas program across all available fiscal years. These queries provided the initial set of project records and the funded publications. Project serial numbers identified from the initial award records were then used in a second set of RePORTER queries to capture related award records that were not returned by the opportunity-number searches. The second-stage queries expanded the award record, identifying sets of supplemental funding opportunity records (e.g., PA-{##}-{###}) for each program, which did not materially change the NIH-funded publication set. RePORTER data for all three programs were downloaded in CSV format on February 5, 2026.

**Table 2. Record for NIH-funded atlas program opportunity number, project serial number, award records.**

| NIH Program | HTAN | HuBMAP | SenNet |
|---|---|---|---|
| Opportunity Numbers | 6 | 9 | 6 |
| Project Serial Numbers | 21 | 45 | 26 |
| Award Records | 203 | 326 | 356 |

The NIH Core Project Number and Full Project Number were treated as distinct project identifiers. The Core Project Number identifies a research project independently of a specific support year and was used as the principal identifier for linking projects to publications and CCs. The Full Project Number represents the complete grant or project identifier. Thus, multiple Full Project Numbers may correspond to the same Core Project Number across support years, application types, or supplements, allowing individual award records to be grouped under a common underlying research project.

## Publication data

Publication records for the three atlas programs were obtained from NIH RePORTER. RePORTER links each publication, identified by its PubMed identifier (PMID), to an NIH Core Project Number. We used these Core Project Numbers to associate publications with the corresponding funded projects and CCs. Because Core Project Numbers represent the overall project rather than a specific budget period, publications could not be assigned to a specific project year or fiscal year of funding. For comparisons across programs with different publication start dates, publication years were aligned relative to the first observed publication year for each atlas program.

Preprint records were matched to their corresponding journal publications to remove duplicate versions of the same work. Candidate preprint-journal publication pairs and other duplicate versions were identified using publication titles and author lists and were manually reviewed before exclusion. Following validation, 886 preprint-journal publication pairs were identified. For each matched pair, the journal publication was retained as the version of record and the corresponding preprint was excluded. Finally, there are 1,255 unique publications, including 1,102 journal publications and 153 preprints. Additionally, 107 publications acknowledge funding by more than one of the three atlas programs.

## Expert data

Researcher names were obtained from NIH award records and publication author lists. Principal investigators (PIs) were identified from the RePORTER Contact PI/Project Leader and Other PI or Project Leader(s) fields. NIH RePORTER defines the PI fields as investigators designated by the organization to direct the research project. For each program, full author names were matched to the PI records, which were categorized as PIs; all other individual authors were classified as researchers. Group authors, including consortium names, were identified separately and excluded from all analyses.

Award and publication name records were combined before disambiguation to enable linkage of individuals across projects, publications, and networks. Names were disambiguated in OpenRefine using iterative application of key-collision and nearest-neighbour clustering algorithms followed by manual review of candidate matches; transformation histories were retained for reproducibility (Ham 2013; OpenRefine 2026). During this process, 58 duplicate investigator records were removed and 566 publication-author names were merged, yielding 8,952 unique individual names, including 215 PIs, across the three atlas programs. A total of 63 group-author records were identified separately, including seven additional records identified during name disambiguation. PI records from all award years were retained and linked to their respective core projects. Following disambiguation, PI-project associations were compared with publication authorship records; the absence of a matched PI authorship was treated as a limitation of the observed publication dataset rather than evidence of absent research activity. Six core projects (five in HTAN and one in HuBMAP) had no funded publications, involving 18 unique PIs. Discussions with NIH program officers suggested that some investigators had left their projects early, while others were associated with newer projects that had not yet reported publications.

### Citation data

The Relative Citation Ratio (RCR) is a publication-level bibliometric metric developed by the NIH Office of Portfolio Analysis to assess the citation influence of scientific publications, accounting for differences across research fields and publication years (Hutchins et al. 2016). By providing a field- and time-normalized measure of citation influence, RCR facilitates comparisons across publications, with an RCR of 1.0 representing the median citation rate of NIH-funded publications in the corresponding field and publication year.

RCR values were obtained from NIH RePORTER on February 5, 2026, and citing-publication data were retrieved from OpenAlex (Priem et al. 2022) on September 24, 2026 using the validated atlas-program publications as focal records. For each focal publication, citing works were linked using publication identifiers and deduplicated so that each atlas program-focal publication-citing work combination represented one citation relationship. The citation dataset retained citing-publication year, journal, OpenAlex field and domain, and institutional country codes, together with the atlas program and collaboration classification of the focal publication. RCR values were available for 807 of 1,447 program-publication records (55.8%), corresponding to 750 of 1,341 unique PMIDs (55.9%). Among the focal publications, *Comprehensive Integration of Single-Cell Data*(Stuart et al. 2019) had the highest citations (16,670). The highest RCR (526.05) was observed for *Integrated analysis of multimodal single-cell data*(Hao et al. 2021).

## Internal collaboration classification and measurement

Internal collaboration (IC) was classified separately within each atlas program according to the number and CC distribution of Core Project Numbers associated with each publication. Single-project publications were linked to one Core Project Number; intra-CC collaborative publications were linked to multiple Core Project Numbers within a single CC; and inter-CCs collaborative publications were linked to multiple Core Project Numbers spanning two or more CCs. Intra- and inter-CCs publications were collectively classified as IC publications, whereas publications linked to a single funded project were classified as single-project publications. For publications associated with more than one atlas program, collaboration type was assigned independently within each program according to its program-specific project and CC associations.

To characterize the organizational form of IC within CCs, we calculated a CC-level Internal Collaboration Type (ICT) score:

$$ICT_d = \frac{N_{Inter,d} - N_{Intra,d}}{N_{Inter,d} + N_{Intra,d}}$$

where $N_{Inter,d}$ and $N_{Intra,d}$ are the numbers of distinct inter-CCs and intra-CC publications associated with CC $d$, respectively. The score ranges from -1 to +1. A score of -1 indicates that all IC publications associated with the CC were intra-CC collaborations, +1 indicates that all were inter-CCs collaborations, and 0 indicates equal numbers of the two collaboration types. The magnitude and direction of the score therefore describe the observed composition of IC publications, rather than the probability that a project will collaborate.

Author-level collaboration type was summarized using an author-level ICT score calculated from each author's IC publications:

$$ICT_a = \frac{N_{Inter,a} - N_{Intra,a}}{N_{Inter,a} + N_{Intra,a}}$$

where $N_{Inter,a}$ and $N_{Intra,a}$ are the numbers of distinct inter-CCs and intra-CC publications associated with author $a$, respectively. Author ICT scores have the same range and interpretation as CC ICT scoring.

## Author participation and co-authorship network construction

Author participation in IC was assessed relative to the complete baseline co-authorship network. Authors with at least one IC publication were classified as IC participants, and participation rates were calculated overall and separately for principal investigators (PIs) and other researchers within each atlas program. IC publication activity was further classified into five ordinal levels based on the number of IC publications per author: highly active (≥10), very active (5-9), moderately active (2-4), slightly active (1), and inactive (0), with distributions compared across atlas programs and author roles.

Program-specific weighted co-authorship networks were constructed from publication authorship data, where nodes represented individual authors and edges connected authors who co-authored one or more publications. Edge weights represented the number of publications co-authored by each author pair. Group authors were excluded. For each atlas program, we generated a baseline network containing all retained publications and an IC network containing only intra- and inter-CCs collaborative publications. For visualization, networks were filtered to retain recurrent co-authorship relationships by removing edges with weights below two (i.e., author pairs with fewer than two co-authored publications) and excluding resulting isolated nodes.

Node size was mapped to cumulative publication count, while node color represented author role. Network layouts were generated using the ForceAtlas2 algorithm (Jacomy et al. 2014) in Gephi (version 0.10.1), with minor manual adjustments to the positions of small network components and node labels to improve visual clarity. Community detection was performed separately using the weighted Louvain algorithm (Blondel et al. 2008) implemented through the cluster_louvain() function in the R igraph package (Csardi and Nepusz 2006).

To examine the association between researcher roles and collaboration patterns, we analyzed 18,679 author-program-publication records representing 10,229 author-program observations. Researchers were classified as PIs or other researchers, and publications were grouped into intra- and inter-CCs collaborations, and single-project. We used three pairwise logistic regression models to compare inter-CCs with single-project publications, intra-CC with single-project publications, and inter-CCs with intra-CC publications. All models included consortium and publication-year fixed effects and adjusted for researcher publication activity using log(1 + total program-associated publications). Comparisons involving intra-CC collaboration were restricted to HuBMAP and SenNet because HTAN had no intra-CC publications.

Adjusted odds ratios (ORs) and 95% confidence intervals were estimated using two-way cluster-robust standard errors at the researcher and publication levels to account for repeated observations. Statistical significance was assessed using two-sided Wald tests, with p-values adjusted across the three comparisons using the Holm method (Holm 1979). Holm-adjusted p-values below 0.05 were considered significant. As a robustness check, we used modified Poisson regression with researcher-clustered standard errors to estimate relative risks (RRs) (Zou 2004) of participation in each collaboration category at the author-program level, adjusting for consortium and publication activity. Additional consortium-specific analyses examined whether the association between PI status and inter-CCs participation was observed across the three programs.

## Citation analysis

Citation influence was first compared using RCR distributions between IC and single-project publications and, within the IC subset, between intra- and inter-CCs publications, stratified by atlas program and CCs. Analyses emphasized median RCR because of the right-skewed RCR distribution. Matched preprints and the most recent publication year in each program were excluded, and each publication contributed only once to a given program-CCs comparison. RCR comparisons were restricted to publications with available RCR values.

Publication citation trajectories were then calculated using the citation window, defined as the difference between the publication years of the citing and focal publications. The disciplinary reach of citations was characterized using the OpenAlex field and domain assigned to each citing publication. Distinct citing works were counted by field, citation window, atlas program and focal-publication collaboration type, and field-by-citation-window distributions were visualized using log-transformed heatmaps. These analyses were used to compare the disciplinary profiles of citations to intra- and inter-CCs, and single-project publications across HTAN, HuBMAP and SenNet.

To examine the association between collaboration patterns and citation performance, publications were classified as intra- and inter-CCs collaboration, and single-project publications. Cumulative and annual new citations were measured across five complete post-publication years and transformed using log(1 + citations) to account for zero counts and reduce the influence of highly cited publications. Pairwise comparisons were conducted between publications from the same consortium and publication year, retaining only strata containing both comparison groups. Differences in mean log-transformed citations were combined using the following weights,

$$w = (n_0 \times n_1)/(n_0 + n_1)$$

where $n_0$ and $n_1$ represent the numbers of publication records in each group within a stratum. The pooled estimates were converted into adjusted ratios of geometric means of (1 + citations) using the exponential function.

Statistical significance was assessed using two-sided stratified permutation tests (25,000 permutations) (Winkler et al. 2015), with collaboration labels permuted within consortium-publication-year strata. We calculated 95% confidence intervals using 2,500 stratified bootstrap resamples and adjusted p-values using the Benjamini-Hochberg false discovery rate (FDR) procedure (Benjamini and Hochberg 1995), applied separately to 14 valid comparisons for each citation outcome. FDR-adjusted q-values below 0.05 were considered significant, and comparisons with fewer than five eligible records in either group were excluded. To assess citation growth over time, we additionally compared slopes of log-transformed annual citations within fixed follow-up cohorts using the same stratified permutation approach.

# Results

## Program organization and funding

**Fig. 1** presents distinct project structures and cost profiles across the three atlas programs. HuBMAP had the largest numbers of unique core project numbers and full project numbers but the lowest aggregate total cost across the included fiscal years. In contrast, HTAN had the lowest number of core and full project numbers but the highest aggregate total cost. SenNet had the largest numbers of RePORTER project records and subprojects, together with the second-highest total cost. Thus, differences in project-record counts across the programs did not parallel differences in total cost.

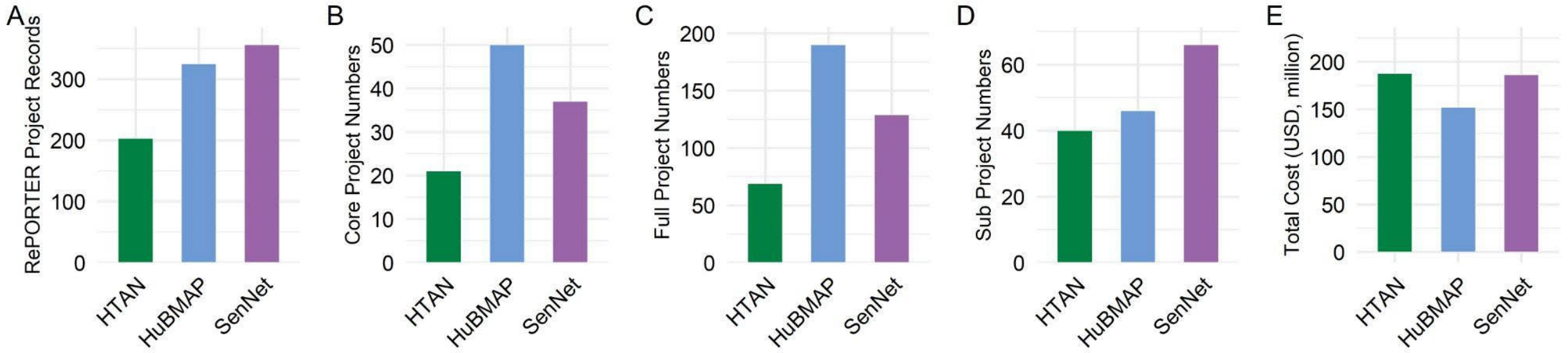

**Figure 1. Project records, project identifiers and total costs across NIH atlas programs. A** Number of NIH RePORTER project records included for HTAN, HuBMAP and SenNet. **B** Number of unique core project numbers. **C** Number of unique project numbers. **D** Number of subprojects. **E** Aggregate total cost across all included fiscal years, expressed in millions of US dollars. Note that three atlas programs have different funding timelines: HTAN from 2018 to 2029, HuBMAP from 2018 to 2025, and SenNet from 2021 to 2026.

Atlas program CC organizations have similar structures with distinct funding patterns. Each program incorporates a data administration center and one or more tissue atlas centers; and both HuBMAP and SenNet incorporate technical development CCs[1](“Expired RFA-CA-21-037: 3D Technologies to Accelerate HTAN Atlas Building Efforts (UH2 Clinical Trial Not Allowed)” n.d.). HuBMAP has the most complex organizational structure, with five CCs, which includes rapid technical integration CC (“Expired RFA-RM-19-002: The Human Biomolecular Atlas Program (HuBMAP): Rapid Implementation of Technologies that Will Accelerate Development of a Framework for Mapping the Human Body at High Resolution (UH3 Clinical Trial Not Allowed)” n.d.) and data demonstration CC that funded projects which make creative uses of HuBMAP data and resources(“Expired RFA-RM-21-027: The Human BioMolecular Atlas Program (HuBMAP) Demonstration Project (U01 Clinical Trial Not Allowed)” n.d.).

## Collaboration patterns across consortium components

**Fig. 2** shows IC across the three atlas programs, with the lowest levels observed in HTAN, intermediate levels in HuBMAP, and the highest levels in SenNet across publication output, core-project involvement, and PI participation. IC publications represented a relatively small share of publication output but involved a large proportion of core projects and PIs, particularly in HuBMAP and SenNet.

**Fig. 3** presents the trends between IC publications and overall publications over time. As RePORTER publications cannot be assigned to specific award or funding years, temporal trends were based on publication year and aligned to each program’s first observed publication. Overall publication output generally increased over the observed publication years. HTAN exhibited relatively low IC publication counts, whereas HuBMAP showed a more substantial increase before its IC output largely stabilized in later observed years. SenNet demonstrated the most pronounced early expansion, with both overall and IC publication counts increasing rapidly during its first five observed publication years.

[1] A technical development division, titled 3D Technologies to Accelerate HTAN Atlas Building Efforts (UH2 Clinical Trial Not Allowed), was twice proposed for HTAN under RFA-CA-20-042 and RFA-CA-21-037 (National Institutes of Health (NIH) 2021a, 2021b).

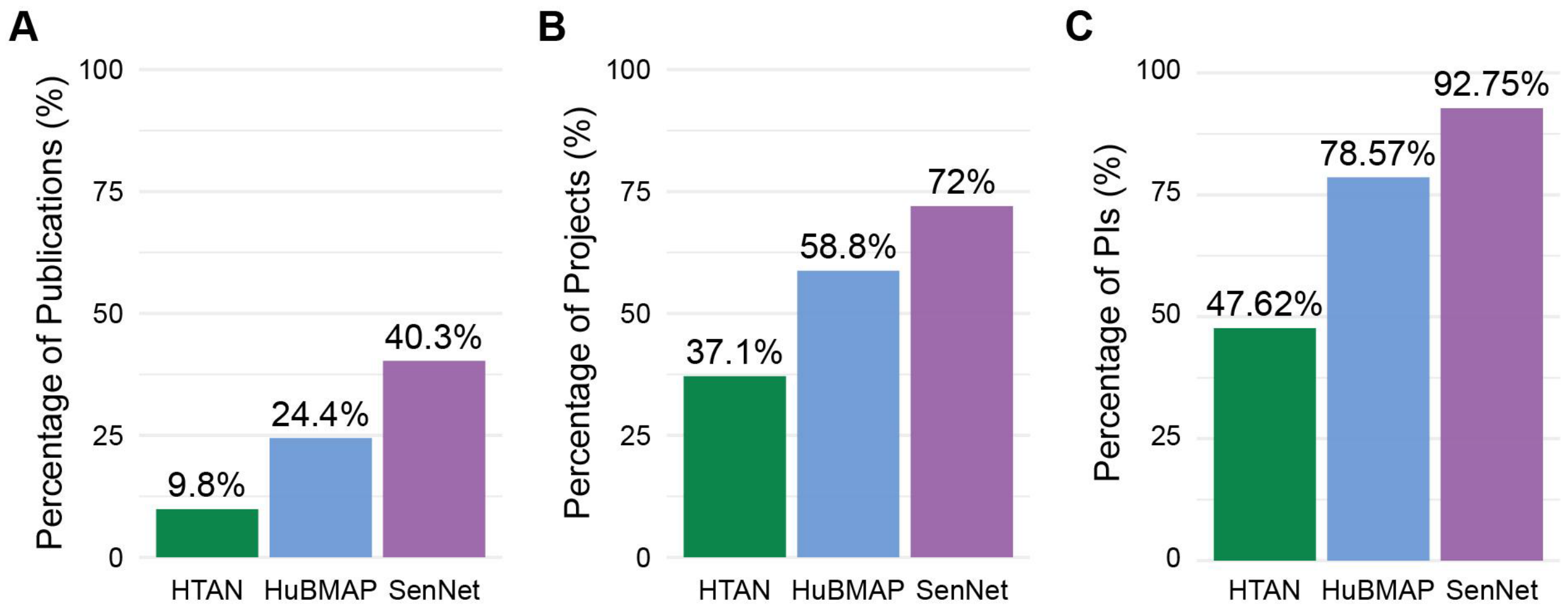

**Figure 2. Internal collaboration across HTAN, HuBMAP, and SenNet. A** Percentage of publications classified as IC publications within each program. **B** Percentage of unique core projects associated with at least one IC publication. **C** Percentage of PIs associated with at least one IC publication.

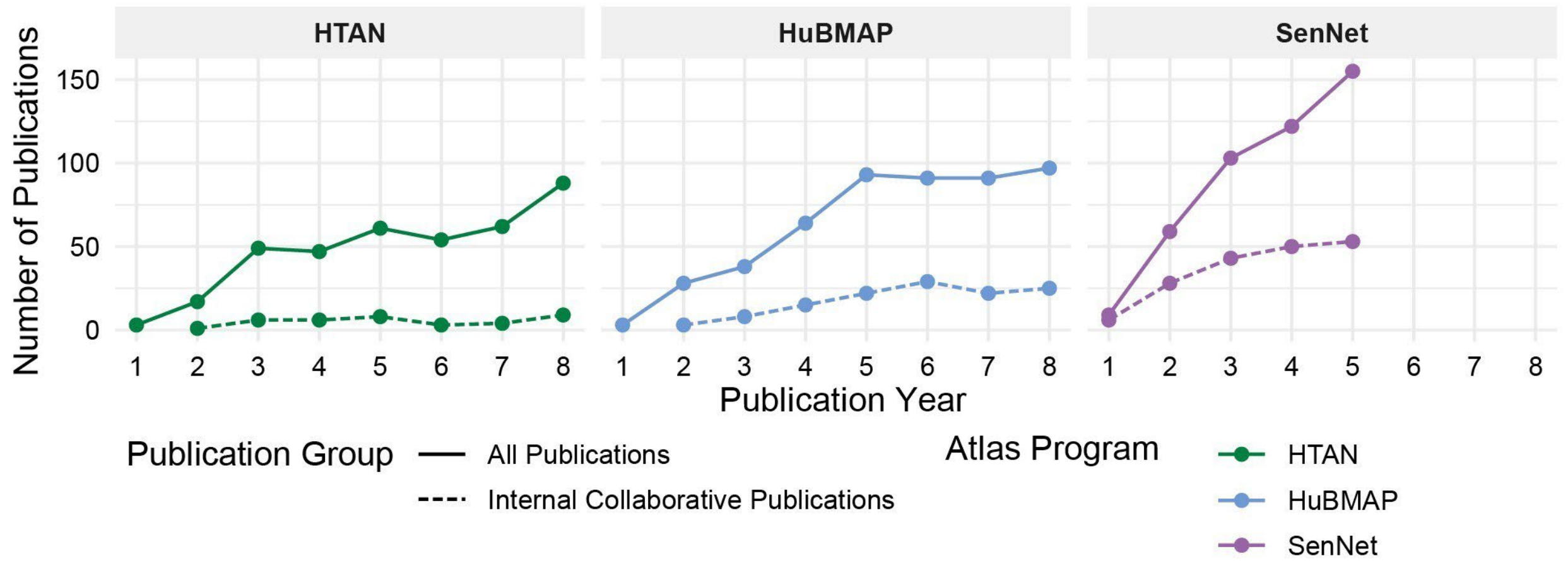

**Figure 3. Temporal trends in publications and internal collaboration across three atlas programs.** Solid lines indicate all publication records, and dashed lines indicate IC publications.

IC patterns varied across CCs and atlas programs. HTAN CCs predominantly engaged in inter-CCs collaboration, whereas HuBMAP and SenNet showed greater variation between intra- and inter-CCs collaboration (**Fig. 4A**). Technology-oriented CCs, particularly TTD and TDA, tended toward intra-CC collaboration (Pattern A), while experimental-oriented CCs, such as HTA and PCA, favored inter-CCs collaboration with CCs of the same functional type (Pattern B) (**Fig. 4B**). However, some CCs showed different preferences, with RTI favoring cross-functional collaboration (Pattern C) and others exhibiting mixed patterns.

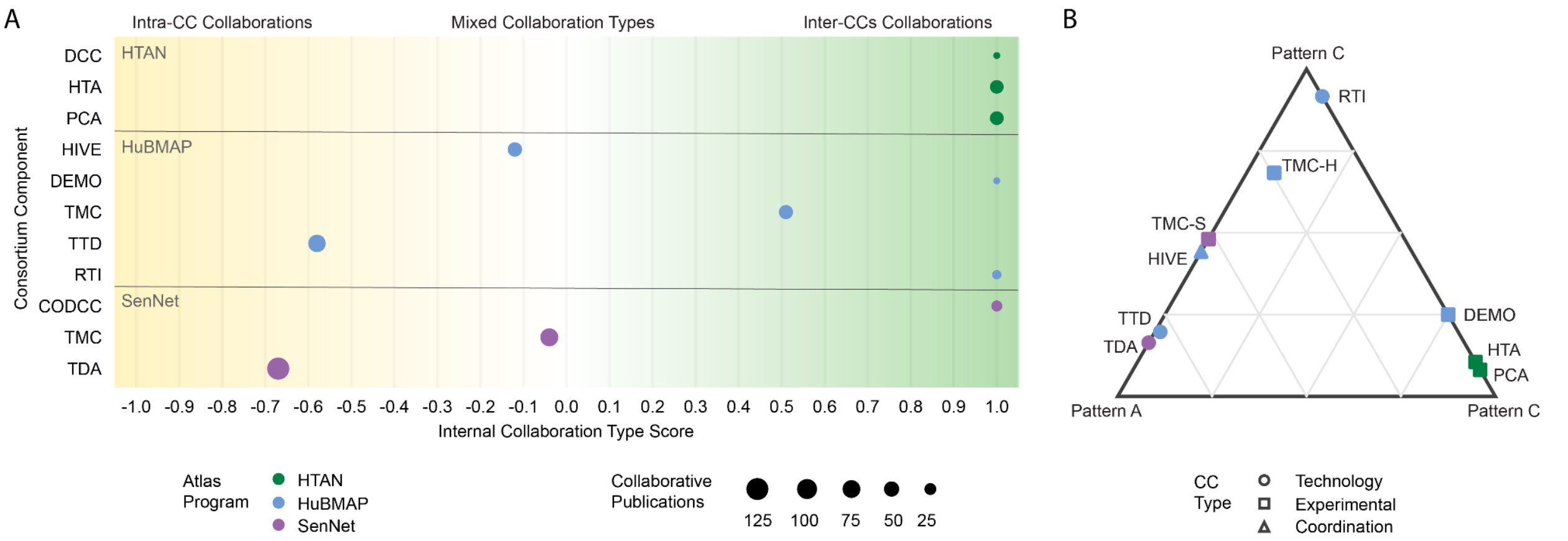

**Figure 4. Collaboration patterns. A** Internal collaboration type across NIH-funded atlas program CCs. Point size represents the number of IC publications associated with each CC, and colour denotes the atlas program. **B** Distribution of CCs in different collaboration patterns. Each point represents one consortium component within an atlas program. Its position reflects the relative shares of pattern A, B, and C, with points closer to a vertex indicating a larger share of that pattern.

# Researcher roles in collaboration

The number of authors were similar across the three atlas programs, but participation in IC differed substantially (**Table 3**). IC publications involved 10.6% of HTAN authors, 24.7% of HuBMAP authors and 44.2% of SenNet authors. Across all three programs, PIs participated in IC publications substantially more frequently than other researchers, with PI participation ranging from 47.6% in HTAN to 92.8% in SenNet, compared with 9.9-43.2% among non-PI researchers.

**Table 3. Atlas program authors and participation in internal collaborative research.**

| Atlas Program | Count of Authors | Count of IC Authors | Percent of Authors in IC publications | Count of PI Authors | Count of PI IC Authors | Percent of PI in IC publications | Percent of Other researcher in IC publications |
|---|---|---|---|---|---|---|---|
| HTAN | 3456 | 366 | 10.59 | 63 | 30 | 47.62 | 9.9 |
| HuBMAP | 3520 | 871 | 24.74 | 70 | 55 | 54.79 | 23.65 |
| SenNet | 3253 | 1439 | 44.24 | 69 | 64 | 87.67 | 43.18 |

Differences in participation were accompanied by differences in IC publication activity (**Table 4**). Most non-PI researchers were inactive, although the inactive fraction decreased from HTAN to HuBMAP to SenNet. PI activity showed a stronger shift across programs: more than half of HTAN PIs had no IC publications, whereas inactivity was uncommon among SenNet PIs. Conversely, the proportion of PIs contributing to at least two IC publications increased from approximately one-quarter in HTAN to more than half in HuBMAP and more than two-thirds in SenNet. Thus, PIs differed from other researchers not only in the likelihood of participating in IC but also in the intensity of their participation.

**Table 4. Distribution of atlas program internal collaboration activity level by author role.**

| Atlas Program | Internal Collaboration Activity Level | Percent of PIs | Percent of Researchers |
|---|---|---|---|
| HTAN | Highly Active (10+) | 3.17 | 0.06 |
| HTAN | Very Active (5-9) | 0.00 | 0.32 |
| HTAN | Moderately Active (2-4) | 22.22 | 1.56 |
| HTAN | Slightly Active (1) | 22.22 | 7.96 |
| HTAN | Inactive (0) | 52.38 | 90.10 |
| HuBMAP | Highly Active (10+) | 5.71 | 0.06 |
| HuBMAP | Very Active (5-9) | 18.57 | 0.90 |
| HuBMAP | Moderately Active (2-4) | 30.00 | 6.14 |
| HuBMAP | Slightly Active (1) | 24.29 | 16.55 |
| HuBMAP | Inactive (0) | 21.43 | 76.35 |
| SenNet | Highly Active (10+) | 18.84 | 0.22 |
| SenNet | Very Active (5-9) | 17.39 | 0.97 |
| SenNet | Moderately Active (2-4) | 33.33 | 10.99 |
| SenNet | Slightly Active (1) | 23.19 | 31.00 |
| SenNet | Inactive (0) | 7.25 | 56.82 |

Author-level network structure and collaboration type further distinguished the programs (**Fig. 5** with PI-labeled networks available at https://cns-iu.github.io/nih-atlas-supporting-information/index.html#co-authorship-network). In the filtered co-authorship networks (retaining only author pairs with at least two co-authored publications), most displayed authors belonged to a dominant connected component, with dense local co-authorship clusters and smaller peripheral components; HuBMAP showed greater fragmentation into secondary components than HTAN or SenNet. ICT-score distributions showed an additional role-dependent pattern. HTAN IC authors were exclusively associated with inter-CCs collaboration, whereas HuBMAP and SenNet spanned both intra- and inter-CCs collaboration. In the latter two programs, non-PI researchers were concentrated toward intra-CC collaboration, whereas PIs were more strongly represented toward the inter-CCs end of the ICT distribution. Together, these results indicate that program differences in internal collaboration encompassed not only the number and activity of participating authors but also the organization and type of collaboration within co-authorship networks.

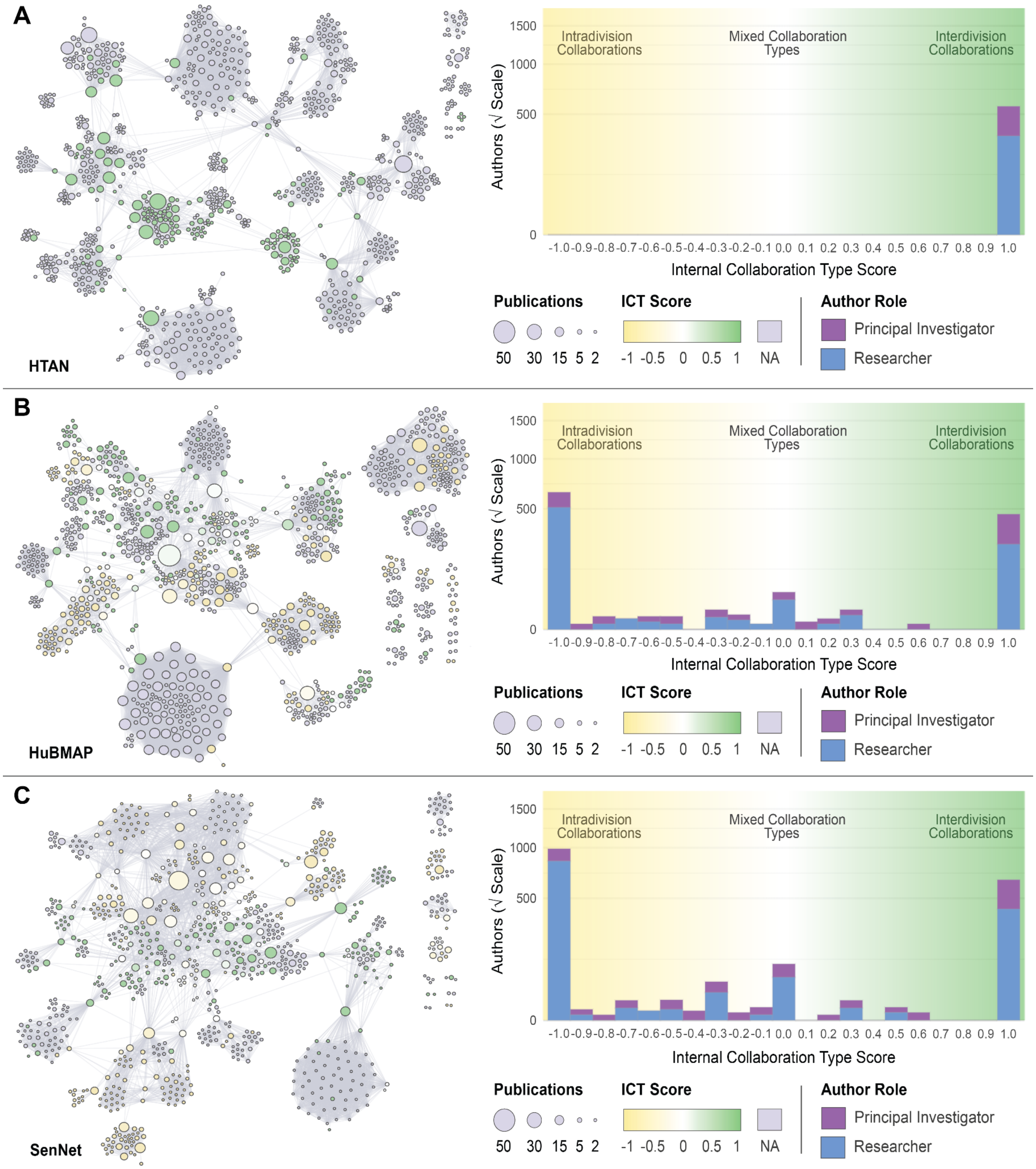

**Figure 5. Co-authorship networks and author-level internal collaboration across NIH atlas programs.** Co-authorship networks and corresponding ICT-score distributions for HTAN (**A**), HuBMAP (**B**) and SenNet (**C**). In the networks, nodes represent authors, edges represent co-authorship, node size reflects publication count and node colour indicates ICT score. Bar charts show ICT-score distributions for PIs and other researchers, with author counts displayed on a square-root scale.

We compared collaboration patterns between PIs and other researchers using logistic regression, adjusting for consortium, publication year, and researcher publication activity (**Table 5**). Inter-CCs publications accounted for 19.0% of PI-associated authorship records, compared with 9.2% for other researchers. After adjusting for consortium, publication year, and researcher publication activity, PIs had significantly higher associations of contributing to inter-CCs rather than single-project publications. A significant association was also observed when inter-CCs collaborations were compared with intra-CC collaborations. In contrast, comparison between intra-CC and single project showed no significant difference. These results indicate that PI involvement is more strongly associated with inter-CCs collaboration than with either intra-CC or single-project publications.

**Table 5. PI status and collaboration patterns.**

| | Model 1<br>Inter-CCs vs<br>single project | Model 2<br>Intra-CC vs<br>single project | Model 3<br>Inter-CCs vs<br>intra-CC |
|---|---|---|---|
| PI status: adjusted OR | 1.72* | 0.95 | 2.57*** |
| 95% CI for OR | [1.14, 2.59] | [0.60, 1.49] | [1.62, 4.07] |
| Logit coefficient, b | 0.540 | -0.054 | 0.944 |
| Cluster-robust SE of b | (0.209) | (0.232) | (0.235) |
| Unadjusted p-value | 0.010 | 0.817 | <0.001 |
| Holm-adjusted p-value | 0.020 | 0.817 | <0.001 |
| Observations (N) | 15,992 | 11,296 | 4,019 |
| PI-associated observations | 1,534 | 1,019 | 595 |
| Other-researcher observations | 14,458 | 10,277 | 3,424 |
| Unique researchers | 8,066 | 5,797 | 2,171 |
| Unique publications (PMIDs) | 1,043 | 860 | 298 |
| Outcome = 1 observations | 1,905 | 2,687 | 1,332 |
| Consortia included | HTAN / HuBMAP / SenNet | HuBMAP / SenNet | HuBMAP / SenNet |
| Consortium fixed effects | Yes | Yes | Yes |
| Publication-year fixed effects | Yes | Yes | Yes |
| Publication activity control | Yes | Yes | Yes |

*Note.* Observations are author-program-publication records. In each model, the outcome equals 1 for the first named publication category and 0 for the reference category. Other researchers are the reference for PI status. Adjusted OR = exp(b); SE denotes the two-way cluster-robust standard error of the logit coefficient (clustered by researcher and PMID). Models adjust for consortium, publication year and log(1 + researcher publication count within consortium). Models 2 and 3 exclude HTAN because it has no intra-CC observations. Significance symbols are based on Holm-adjusted p-values across the three pairwise tests: * $p < 0.05$; ** $p < 0.01$; *** $p < 0.001$.

## Citation impact across collaboration patterns

Relative citation influence varied by collaboration status and CCs (**Table 6**). In HTAN, IC publications had higher median RCRs than single-project publications in both HTA (3.08 compared with 2.09) and PCA (3.08 compared with 1.20). HuBMAP showed no consistent IC-associated difference across CCs, whereas SenNet IC publications generally had higher median RCRs than single-project publications. Within IC publications, collaboration type showed further heterogeneity. In HuBMAP, the median RCR was higher for intra-CC publications in HIVE (6.55) than for inter-CCs publications (1.73), whereas TTD showed a higher median RCR for inter-CCs publications (5.02) than for intra-CC publications (2.74). In SenNet, the median RCR was higher for inter-CCs publications in TDA (6.48) than for intra-CC publications (3.27), whereas TMC showed the opposite pattern, with a higher median RCR for intra-CC publications (4.56) than for inter-CCs publications (2.61). Overall, the association between IC and citation influence varied across programs and CCs rather than following a uniform pattern.

We compared atlas programs' publication impact using citations and RCR overall and by IC publication groups, shown in **Table 6**. Overall, HuBMAP has the highest total citation counts (67,000) and mean citations per publication (143) and highest variance compared to the other atlas programs. On average, internal collaboration publications received more citations for the HTAN and SenNet programs, and fewer citations for publications reported by HuBMAP. The RCR provides a normalized citation rate for NIH-funded publications that are benchmarked to a field and time using publication co-citation network analysis. An RCR score of 1 indicates that a publication received the same number of citations per year as other papers within its field, while a paper with a score or 3 has received three times as many citations compared to other publications from its field. HuBMAP has the highest overall RCR of all programs (10.52). Internal collaborative publications also have higher mean RCR scores compared to single-project publications across all atlas programs. The greatest difference between RCR scores for internal collaboration and single-project publications is seen in HuBMAP (2.39); while the SenNet program has the smallest difference in RCR values (0.03).

**Table 6. Atlas program publication impact across collaboration types.**

| Program | Publication Group | Total Citations | Mean Citations | Std. Dev. Citations | Var. Citations | Mean RCR | Std. Dev. RCR | Var. RCR |
|---|---|---|---|---|---|---|---|---|
| HTAN | All Publications | 36294 | 106.12 | 238.04 | 56664.52 | 4.72 | 8.54 | 72.88 |
| HTAN | IC Publications | 5619 | 181.26 | 300.97 | 90583.33 | 5.55 | 6.67 | 44.47 |
| HTAN | Single-project Publications | 30675 | 98.63 | 230.10 | 52944.05 | 4.63 | 8.74 | 76.31 |
| HuBMAP | All Publications | 105996 | 226.49 | 1169.28 | 1367206.00 | 10.52 | 41.81 | 1747.83 |
| HuBMAP | IC Publications | 21670 | 191.77 | 634.02 | 401986.43 | 12.34 | 43.90 | 1926.90 |
| HuBMAP | Single-project Publications | 84326 | 237.54 | 1294.58 | 1675941.17 | 9.95 | 41.20 | 1697.23 |
| SenNet | All Publications | 28145 | 75.66 | 313.90 | 98534.50 | 7.3 | 26.59 | 706.79 |
| SenNet | IC Publications | 12563 | 84.89 | 173.44 | 30082.28 | 7.32 | 10.45 | 109.25 |
| SenNet | Single-project Publications | 15582 | 69.56 | 379.48 | 144005.75 | 7.29 | 34.33 | 1178.38 |

Inter-CCs publications showed significantly higher cumulative citation levels than single-project publications across the first five post-publication years (H1-H5; all FDR-adjusted q<0.05; **Table 7**), after accounting for differences in consortium and publication year. This difference was evident from the first year (adjusted citation ratio=1.57, q=0.021) and remained significant in subsequent years, reaching a ratio of 3.47 at H4 (q=0.020). A similar pattern was observed for annual new citations, with inter-CCs publications showing significantly higher citation levels in each of the first five years (all q=0.020). These results indicate a consistent association between inter-CCs collaboration and higher early citation performance, evident in both cumulative citations and citations received within individual years.

**Table 7. Citation performance between IC and single-project publications.**

| Citation window | Single-project Publications | Inter-CCs publications | Cumulative citations | | Annual new citations | |
|---|---|---|---|---|---|---|
| | | | ratio (95% CI) | p / FDR q | ratio (95% CI) | p / FDR q |
| H1 | 653 | 75 | 1.57* (1.22-2.06) | 0.006 / 0.021 | 1.57* (1.21-2.05) | 0.005 / 0.020 |
| H2 | 483 | 62 | 1.76* (1.34-2.34) | 0.003 / 0.021 | 1.66* (1.23-2.25) | 0.007 / 0.020 |
| H3 | 328 | 42 | 1.94* (1.39-2.68) | 0.005 / 0.021 | 1.9* (1.35-2.68) | 0.006 / 0.020 |
| H4 | 174 | 18 | 3.47* (2.09-5.77) | 0.001 / 0.020 | 3.23* (1.79-5.59) | 0.002 / 0.020 |
| H5 | 84 | 8 | 3.85* (1.68-8.38) | 0.012 / 0.029 | 4.65* (1.92-10.44) | 0.007 / 0.020 |

**Fig. 6** presents distinct disciplinary citation impacts across the three atlas programs. HuBMAP showed the greatest overall citing paper reach across collaboration groups. In HTAN and SenNet, inter-CCs publications attracted more distinct citing papers per focal publication than other publications, with SenNet intra-CC publications showing an intermediate pattern. Life Sciences dominated citing paper reach across all programs, whereas Social Sciences contributed minimally. HuBMAP inter-CCs publications were particularly concentrated in Life Sciences, while its intra-CC and single-project publications received comparatively greater citing paper reach from Health and Physical Sciences.

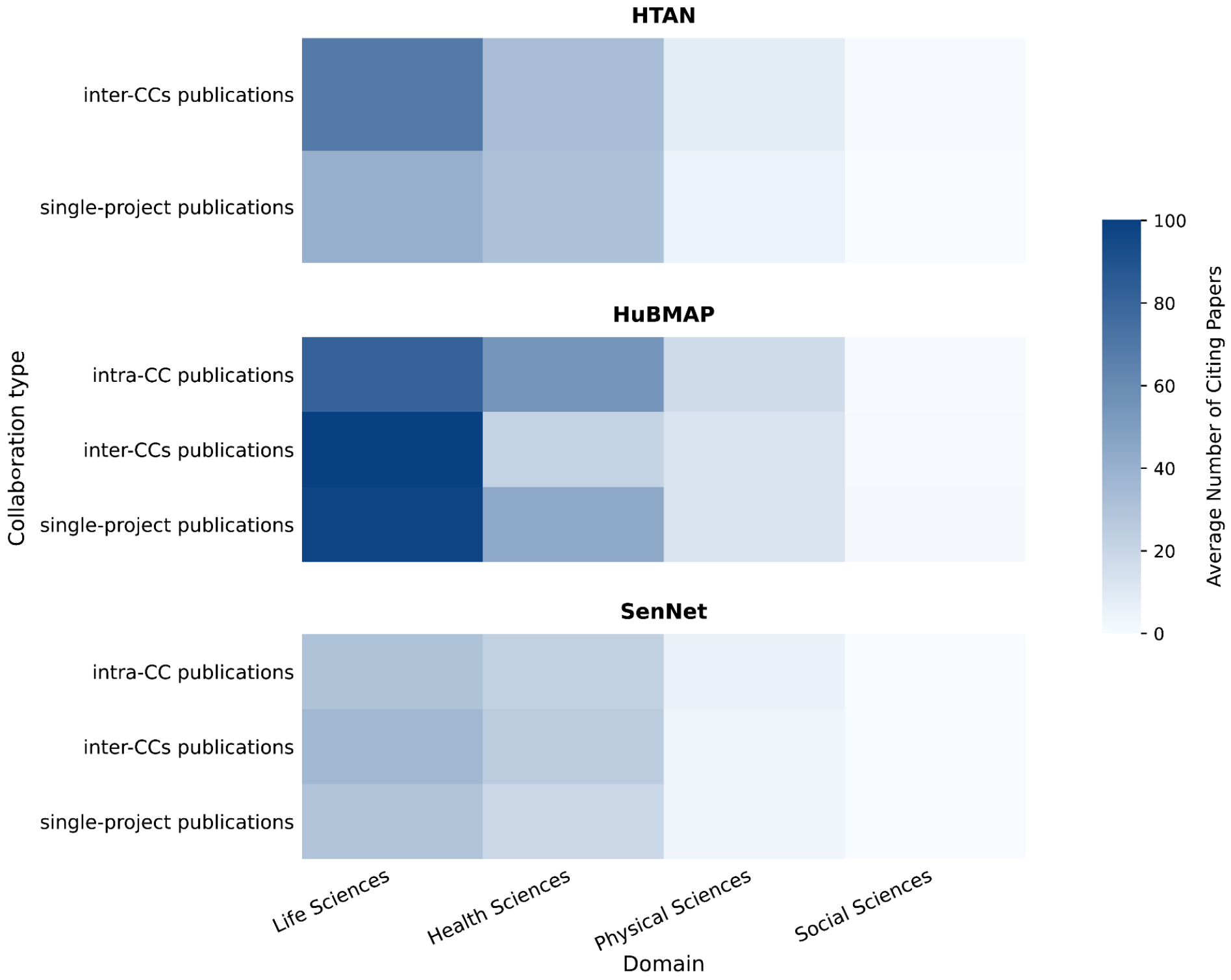

**Figure 6. Disciplinary distribution of citing publications across collaboration types.** Heatmaps show the number of distinct citing publications per funded publication for HTAN, HuBMAP and SenNet, stratified by funded-publication collaboration type and OpenAlex domain of the citing publication. No retained HTAN publications were classified as intra-CC collaborations.

# Discussion and conclusion

By linking NIH-funded projects, publications, researchers, and citation records, this study provides an insight of collaboration within three atlas programs. IC accounted for a minority of publication output but involved many funded projects and PIs, particularly in HuBMAP and SenNet. Thus, the frequency of collaborative publications and the breadth of participation across an initiative's funded portfolio describe different aspects of collaboration. More importantly, distinguishing publications linked to multiple projects within a CC from those spanning CCs of the same or different functional types reveals organizational patterns that would be obscured by institution-level or aggregate publication counts. This extends the evaluation of team science beyond the volume of output to the structure of publication-documented collaboration (Börner et al. 2021; Hall et al. 2018).

Beyond the overall extent of internal collaboration, the differences across CCs highlight the role of functional specialization in shaping collaboration patterns. Technology-oriented CCs, notably TTD and TDA, were more frequently associated with intra-CC collaboration, consistent with research linking collaboration within research units to specialized laboratory activities and the division of labor (Iglič et al. 2017; Haeussler and Sauermann 2020). Experimental-oriented CCs, including HTA and PCA, instead showed greater involvement in inter-CC collaboration among components with the same functions. The production of integrated tissue atlases depends on coordinated sampling, shared experimental protocols, standardized metadata, and data integration across research centers, as demonstrated by the organizational frameworks of HuBMAP and HTAN (HuBMAP Consortium et al. 2019; de Bruijn et al. 2025). This finding is consistent with research emphasizing the importance of shared knowledge and cognitive proximity in establishing collaboration (Koopmann et al. 2021). However, functional similarity alone does not explain the variation observed across CCs, particularly RTI's stronger association with cross-functional collaboration. Previous studies have shown that collaboration is often greatest at intermediate levels of research topic similarity, where shared knowledge is combined with complementary expertise (T. B. Smith et al. 2021). Research on multiteam systems likewise emphasizes coordination between specialized teams with distinct functions (Zaccaro et al. 2020), while longitudinal evidence from a multidisciplinary research center shows that cross-departmental collaboration can increase as research applications diversify (Pepe and Rodriguez 2010).

The functional organization of CCs shapes opportunities for collaboration, while researcher roles help explain who participates across these organizational boundaries. PI status was significantly associated with inter-CCs rather than single-project or intra-CC authorship, even after accounting for publication activity (Table 5). This finding extends previous evidence that senior academics and PIs tend to have higher research productivity and broader collaboration networks (Abramo et al. 2014; Kastrin et al. 2018), by showing that the difference is concentrated in cross-component collaboration rather than general publication activity. The association is consistent with the role of PIs in coordinating research across organizational boundaries, as emphasized in research on PIs as boundary spanners (Mangematin et al. 2014). Studies of NIH-funded research networks similarly found that center leaders were more likely to establish ties with researchers in other roles (The Centers for Population Health and Health Disparities Evaluation Working Group and Okamoto 2015), while PIs increasingly occupied representative and gatekeeper positions between research centers (Ducharme et al. 2024).

In terms of citation impact, the higher citation levels associated with inter-CCs collaboration align with previous evidence that research spanning organizational boundaries can attract greater scientific attention. Publications supported by multiple CTSA hubs exhibited stronger citation influence than single-hub publications (Llewellyn et al. 2020), while larger scientific teams with greater organizational diversity were more likely to produce highly cited research (Yoo et al. 2024). However, the variation in RCR across programs and CCs indicates that the relationship between organizational collaboration and citation influence is not uniform. This is consistent with Fan et al. (2020), who found that citation outcomes differed across institutional collaboration patterns, suggesting that the organizational configuration of collaboration matters beyond the presence of collaboration itself. Furthermore, the higher cumulative and annual citation levels associated with inter-CCs publications were evident from the first post-publication year and persisted across all five analyzed citation windows, reflecting a sustained citation advantage rather than demonstrated faster citation growth.

These findings have practical implications for the funding, management, and impact reporting of future Big Science programs involving hundreds of researchers from different disciplines. For funding agencies, the observed collaboration patterns suggest matching funding structures to research functions:

grouping related technology-development projects within a common CC for specialized technical work, supporting distinct but coordinated experimental CCs for cross-center mapping activities, and establishing dedicated connections across functional types when technology integration is required. For consortium managers, the collaboration patterns observed in these three atlas programs suggest tailoring coordination activities to CC functions. Managers could support joint method development and validation within technology-oriented CCs, promote shared experimental protocols and data standards across experimental CCs, and facilitate collaboration between technology-integration teams and other functional components. Given the stronger association between PI status and inter-CC participation, PIs could be involved in cross-component planning alongside other participating researchers. For impact reporting, combining publication output with the breadth of project participation, normalized citation influence, citation levels over comparable follow-up periods, and the disciplinary origins of citing publications provides a more informative picture of program outcomes than publication or citation totals alone (Hall et al. 2018; Börner et al. 2021). These publication-based indicators should be considered alongside other atlas outputs, including datasets, software, standards, and shared infrastructure (Börner et al. 2025; de Bruijn et al. 2025).

Several limitations should be considered when interpreting these findings. (1) Internal collaboration was inferred from NIH RePORTER project-publication linkages and CC assignments rather than directly observed collaborative activity. Publications classified as single project outputs may still involve substantial collaboration, while project-publication links may be incomplete or incorrectly assigned and cannot be attributed to a specific funding year; funding acknowledgements also do not specify the contribution of individual projects. (2) Comparisons across HTAN, HuBMAP and SenNet are descriptive and should not be interpreted causally because the programs differ in scientific objectives, organizational structure, funding history and observation period. Co-authorship captures publication-documented relationships rather than all forms of scientific interaction, and PI status identifies formal project leadership rather than an individual's specific role in a collaboration. (3) RCR measures field- and time-normalized citation influence rather than research quality itself. RCR values for recently published articles may be unavailable because insufficient citation data have accumulated. Consequently, analyses restricted to publications with available RCR values may underrepresent more recent research. OpenAlex domains also provide a broad, algorithmically assigned representation of disciplinary reach and therefore simplify the multidimensional subject content of citing publications.

# Acknowledgements

We thank Ajay Pillai, Zorina Galis, and Dena Procaccini for their expert comments and suggestions regarding award record reconciliation. We also thank Mirian Ramirez for valuable feedback on an earlier version of the manuscript. K.B. was supported by the following National Institutes of Health (NIH) awards: OT2OD026671, OT2OD033756, U24CA268108, OT2OD030545. The funders had no role in the study design, data collection and analysis, the decision to publish, or preparation of the manuscript. The content is solely the responsibility of the authors and does not necessarily represent the official views of the National Institutes of Health.

# Ethics declarations

## Competing interests

The authors declare no competing interests.